\documentclass[aps,twocolumn,showpacs]{revtex4}
\usepackage{graphics}
\usepackage{graphicx}

\begin{document}
\title{Carrier heating and high-order harmonics generation \\
in doped graphene by a strong ac electric field}
\author{F.T. Vasko}
\email{ftvasko@yahoo.com}
\affiliation{Institute of Semiconductor Physics, NAS of Ukraine,
Pr. Nauki 41, Kiev, 03028, Ukraine }
\date{\today}

\begin{abstract}
The nonlinear response of electrons (holes) in doped graphene on ac
pumping is considered theoretically for the frequency region above the
energy relaxation rate but below the momentum and carrier-carrier
scattering rates. Temporally-dependent heating of electrons by a strong
ac field, which is described within the energy balance approach, leads
to an effective generation of high-order harmonics. The efficiency of
up-conversion of the 1 mm radiation into the third harmonic by a 
single-layer graphene is about $10^{-7}$ at pumping level $\sim 100$ 
kW/cm$^2$, room temperature, and concentration $\sim 5\times 10^{11}$ 
cm$^{-2}$.
\end{abstract}


\maketitle
The nonlinear response of carriers in graphene has been studied during the past
years for the cases of the heating by a strong dc electric field and of the
interband pumping in the visible or IR spectral regions. In the microwave
frequency region, the nonlinear response is also effective because the massless
and gapless energy spectrum and a suppressed relaxation of low-energy carriers.
\cite{1} It is important that different semiconductor structures is widely used
for the multiplication (up-conversion) of the microwave pumping into THz radiation. \cite{2} Recently, the frequency multiplier in the GHz region, 
which is based on the nonlinear response of a bipolar graphene near the electroneutrality point, have been demonstrated. \cite{3} Such a multiplier 
should be effective in the frequency range $\omega\ll\nu_{gr}$ where $\nu_{gr}$ 
is the generation-recombination rate (in short devices, a fast contact 
injection can reduce this frequency restriction). For the high-frequency 
region, the collisionless regime of the high-order harmonics generation was considered in Ref. 4. These calculations are valid under the condition $\omega\gg\nu_m$ which is correspondent to the THz pumping because the 
momentum relaxation rate $\nu_m\sim 10^{13}$ s$^{-1}$ for a doped sample 
with the mean momentum $\overline{p}$. \cite{1} The regime of nonlinear 
response under a strong pumping in the mm frequency range, when 
$\nu_m\gg\omega$, is not considered up to now.

In this paper we examine the heating of carriers and the high-order harmonics generation in doped graphene under a strong ac electric field ${\bf E}_t=
{\bf E}\cos\omega t$ of the mm frequency range, so that $\omega\ll\nu_m$. We restrict ourselves by the weak anisotropy case, $eE/\nu_m\ll\overline{p}$, 
when the electron distribution function $f_{pt}+\Delta f_{{\bf p}t}$ is written through the isotropic part, $f_{pt}$, and the frequency-independent anisotropic addition
\begin{equation}
\Delta f_{{\bf p}t}\simeq\frac{e{\bf E}_t\cdot{\bf p}}{p\nu_p}\left(
-\frac{\partial f_{pt}}{\partial p}\right) ,
\end{equation}
where $\nu_p$ is the momentum relaxation rate. The symmetric part of 
distribution is governed by the averaged over $\bf p$-plane equation (see 
similar calculations for heating by a dc field in Ref. 5)
\begin{equation}
\frac{\partial f_{pt}}{\partial t}+\overline{e{\bf E}_t\cdot\frac{\partial
\Delta f_{{\bf p}t}}{\partial {\bf p}}}=\sum\limits_r J_r \left( f_t|p\right) .
\end{equation}
Here the collision integrals $J_r \left( f_t|p\right)$ describe the nonelastic
relaxation by acoustic phonons ($r=ac$) and the carrier-carrier scattering
($r=cc$). We neglect the relaxation by optical phonons because the energy of hot
carriers is less than the optical phonon energy. The Joule heating contribution
is described by
\begin{equation}
\overline{e{\bf E}_t\cdot\frac{\partial\Delta f_{{\bf p}t}}{\partial{\bf p}}} 
=\frac{(eE)^2}{2p}\frac{\partial}{\partial p}\left[\frac{p}{\nu_p}
\left( -\frac{\partial f_{pt}}{\partial p} \right)\right]\cos^2 \omega t ,
\end{equation}
where a time dependency of $f_{pt}$ appears due to the proportional to
$\cos^2\omega t$ factor.

The concentration of carriers and the current density, $n$ and ${\bf I}_t$, are
determined through $f_{pt}$ and $\Delta f_{{\bf p}t}$ according to the standard
formula
\begin{equation}
\left| \begin{array}{*{20}c} n \\ {\bf I}_t \\ \end{array} \right| =
4\int \frac{d{\bf p}}{(2\pi\hbar )^2}\left|\begin{array}{*{20}c} f_{pt} \\
 e{\bf v}_{\bf p}\Delta f_{{\bf p}t}  \\ \end{array} \right| ,
\end{equation}
where ${\bf v}_{\bf p}=\upsilon {\bf p}/p$ is the velocity of electron with the
momentum $\bf p$ and $\upsilon\simeq 10^8$ cm/s is the characteristic velocity of 
the linear dispersion law. Using Eq. (1) and introducing the nonlinear conductivity $\sigma_t$ according to ${\bf I}_t=\sigma_t{\bf E}_t$, one obtains
\begin{equation}
\sigma_t =\frac{e^2\upsilon}{\pi\hbar^2}\int\limits_0^\infty\frac{dpp}{\nu _p}
\left( -\frac{\partial f_{pt}}{\partial p} \right)\approx\frac{e^2}{\pi\hbar^2}
\frac{\upsilon}{v_d}f_t .
\end{equation}
The right-hand part here is transformed for the case of the short-range scattering, when $\nu_p\simeq v_dp/\hbar$ is written through the characteristic velocity $v_d$; this approach is valid up to energies $\sim$100 meV. \cite{6} 
Here $f_t\equiv f_{p=0t}$ stands for the time-dependent maximal distribution 
at $p=0$.

Because of predominance of the intercarrier scattering in a heavily doped 
graphene if $\omega <\nu_{cc}$, the collision integral $J_{cc}$ imposes the quasiequilibrium distribution $\widetilde{f}_{pt}=f_t/[\exp (\upsilon p/T_t)
(1-f_t)+f_t]$ written through the time-dependent effective temperature, $T_t$, 
and $f_t$. These parameters are related by the normalization condition and 
the energy balance equation. Introducing the integrals over the dimensionless momentum
\begin{eqnarray}
H_s (f_t )=\int\limits_0^\infty\frac{dxx^s}{e^x (1-f_t )+f_t}, ~~~ s=1,2 , \\
K(f_t)=\frac{1-f_t}{f_t}\int\limits_0^\infty\frac{dxx^4e^x}{\left[ e^x(1-f_t)
+f_t \right]^2} , \nonumber
\end{eqnarray}
one transforms the normalization requirement in Eq. (4) as follows
\begin{equation}
n = \frac{2}{\pi }\left(\frac{T_t}{\hbar\upsilon}\right)^2f_t H_1 (f_t ) .
\end{equation}
The balance equation for the energy density, $4\int d{\bf p}f_{pt}\upsilon p
/(2\pi\hbar )^2$, is transformed into
\begin{eqnarray}
\frac{\partial }{\partial t}\left[ T_t\frac{2}{\pi}\left(\frac{T_t}{\hbar\upsilon}
\right)^2f_t H_2(f_t )\right] -\sigma_t E_t^2 \nonumber \\
=\frac{v_e}{\upsilon}\frac{2}{\pi}\left(\frac{T_t}{\hbar\upsilon}\right)^2
\frac{T_t^2}{\hbar}\left( 1-\frac{T_t}{T_{ph}} \right) K(f_t ) ,
\end{eqnarray}
where the Joule contribution was written through the time-dependent conductivity
$\sigma_t\propto f_t$, see Eq. (5). The energy relaxation contribution in the
right-hand side of (8) is written for the case of the quasielastic scattering by
acoustic phonons at temperature $T_{ph}$. \cite{5} Here we used the energy relaxation rate $v_ep/\hbar$ written through the characteristic velocity $v_e\ll\upsilon , v_d$.

Numerical solution of Eq. (8) is performed under the relation (7) and the periodicity conditions $T_{t+\pi /\omega}=T_t$ and $f_{t+\pi /\omega}=f_t$. 
The nonequilibrium distribution $\widetilde{f}_{pt}$ is determined through 
$T_t=T+\delta T_t$ and $f_t=f+\delta f_t$. Here we separated the time-averaged parts, $T$ and $f$, which describe hot electrons, and the oscillating contributions, $\delta T_t$ and $\delta f_t$. The calculations are performed 
at $T_{ph}=$300 K for the concentrations $n=5\times 10^{11}$ cm$^{-2}$ and 
$10^{12}$ cm$^{-2}$, the intensities up to 200 kW/cm$^2$, and the characteristic velocities $v_d\simeq 3\times 10^7$ cm/s and $v_e\simeq$51.4 cm/s. The
dependencies of time-averaged contributions $T$ and $f$ on the pumping 
intensity $S$ are shown in Figs. 1a and 1b. At low intencities of pumping 
$S<$5 kW/cm$^2$, $T_t$ and $f_t$ vary strongly in analogy with to the results on 
heating by a dc electric field. \cite{5} A slowness of these dependencies
takes place at higer pumping levels, in the region $S\sim$10 - 100 kW/cm$^2$.
Since the relaxation rates increase with energy, both $T_t$ and $f_t$ increase with concentration. Notice, that a response on a probe dc field (photoconductivity) is described by Eq. (5) with the time-averaged $f$ so 
that Fig. 1b determines the photoconductivity versus $S$.
\begin{figure}[ht]
\begin{center}
\includegraphics{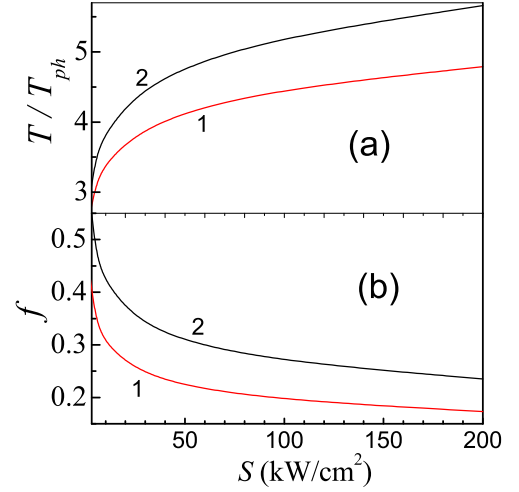}
\end{center}\addvspace{-1 cm}
\caption{Time-averaged temperature $T$ (a) and maximal distribution $f$ (b)
versus pump intensity $S$ plotted for concentrations $5\times 10^{11}$ 
cm$^{-2}$ (1) and $10^{12}$ cm$^{-2}$ (2). }
\end{figure}
\begin{figure}[ht]
\begin{center}
\includegraphics{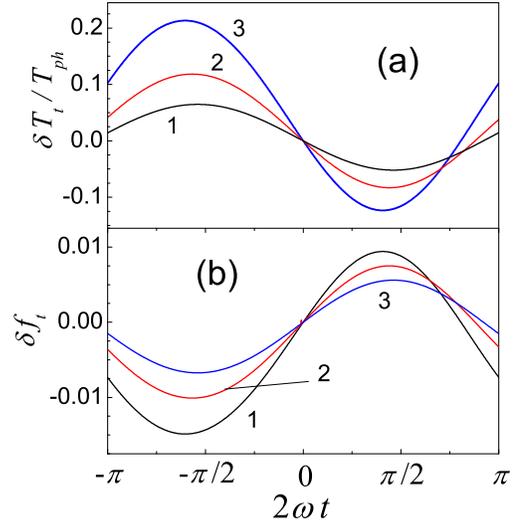}
\end{center}\addvspace{-1 cm}
\caption{Nonharmonic time-dependent contributions to temperature $\delta T_t$
(a) and to maximal distribution $\delta f_t$ (b) for the concentration
$5\times 10^{11}$ cm$^{-2}$ and the 1 mm pumping intensities 25 kW/cm$^2$ (1), 
50 kW/cm$^2$ (3), and 100 kW/cm$^2$ (3). }
\end{figure}

The time-dependent contributions $\delta T_t=T_t -T_{t=0}$ and $\delta f_t=
f_t -f_{t=0}$ are shown in Figs. 2a and 2b for the region $S=$25 - 100 kW/cm$^2$
at concentration $5\times 10^{11}$ cm$^{-2}$. In particular, $\delta f_t$ determines the time-dependent part of $\sigma_t$, i. e. the nonlinear current under a strong pumping. Similar dependencies for 
the concentration $10^{12}$ cm$^{-2}$ are not shown.

Further, we discuss the efficiency of transformation of the millimeter pumping
into THz signal for the case of normal propagation both the pumping wave and 
the $k$th order response determined by the field ${\bf E}_z\exp (-ik\omega t)$.
Substituting $\widetilde{f}_{pt}$ into the 2D current density of Eq. (4) one obtains the Fourier expansion ${\bf I}_t=\sum\nolimits_k{\bf I}_{k\omega}\exp 
(-ik\omega t)$ with the nonzero odd harmonics (it is due to in-plane isotropy 
of the problem). The $k$th harmonic of in-plane THz field, ${\bf E}_z$, is governed by the wave equation: \cite{7}
\begin{equation}
\frac{d^2{\bf E}_z}{dz^2}+\epsilon\left(\frac{k\omega}{c}\right)^2
{\bf E}_z +i\frac{4\pi}{c^2}k\omega{\bf I}_{k\omega }\delta_N (z) = 0 ,
\end{equation}
where $\epsilon$ is the uniform dielectric permittivity and $\delta_N (z)$ is 
the transverse form-factor of the nonlinear current calculated above. For the 
$N$-layer graphene stack, this factor is normalized as $\int_{-d/2}^{d/2}dz\delta_N(z)=N$ with $d\to 0$. The solution of Eq. (9) at $z\neq 0$ takes 
form ${\bf E}_z={\bf E}_{k\omega}\exp (i\kappa_{k\omega}|z|)$ with the wave 
vector $\kappa_{k\omega}=\sqrt{\epsilon}k\omega /c$. The amplitude of $k$th harmonic, ${\bf E}_{k\omega}=-N{\bf I}_{k\omega}(2\pi /\sqrt{\epsilon}c)$ is obtained from the boundary condition at $z\to 0$.
\begin{figure}[ht]
\begin{center}
\includegraphics{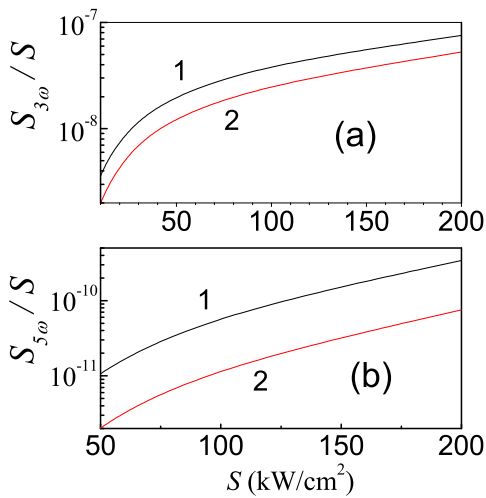}
\end{center}\addvspace{-1 cm}
\caption{Effeciency of transformation into $k$th harmonics, $S_{k\omega } /S$,
versus pumping intensity $S$ for $k=3$ (a) and $k=5$ (b) at doping levels 
$5\times 10^{11}$ cm$^{-2}$ (1) and $10^{12}$ cm$^{-2}$ (2). }
\end{figure}

The efficiency of up-conversion of 1 mm pumping of intensity $S$ into THz signal 
of intensity $S_{k\omega }$ is determined by the ratio $S_{k\omega }/S=
4|{\bf E}_{k\omega}|^2/E^2$. Substituting ${\bf I}_{k\omega }$ from Eq. (4) 
we plot this efficiency for $k=$3 and 5, see Figs. 3a and 3b, where third
and fifth harmonics are correspondent to 0.9 THz and 1.5 THz signals, respectively. 
One obtains the output intensities $S_{3\omega}\sim$1.1 - 8.9 mW/cm$^2$ for the 
pumping range 50 - 150 kW/cm$^2$ at concentration $5\times 10^{11}$ cm$^{-2}$ 
while $S_{5\omega}$ corresponds up to $\sim$10 $\mu$W/cm$^2$ range. Since 
$S_{k\omega}\propto N^2$ for the $N$th layer structure of epitaxial graphene, \cite{8} 
the up-converted 0.9 THz signal appears to be about 0.1 W/cm$^2$ (or $\sim$0.2 
mW/cm$^2$ at 1.5 THz) under the 100 kW/cm$^2$ pumping at the $5\times 10^{11}$ 
cm$^{-2}$ doping level and $N=$5.

Next we briefly discuss the assumptions used in our calculations. We describe the
heating of carriers with the use of the quasiequilibrium distribution approach.
This is valid for the heavily doped graphene which is suitable for an efficient
up-conversion. Typically, $\nu_m\sim\nu_{cc}$ in doped samples and, together with
the low-frequency condition $\omega <\nu_m$, we arrive at the restriction
$\nu_e <\omega <\nu_m,~\nu_{cc}$ which is correspondent to the $10^9$ s$^{-1}$ -
$10^{13}$ s$^{-1}$ frequency region. In spite of $f_{p=0t}<1$ we neglected
the interband generation-recombination processes because the density of states 
vanishes at low energies. Also, the interband optical transitions are not essential in the 
spectral region up to 5$\omega$. Beside of this, we perform 
the simplified electrodynamical estimates of the efficiency of up-conversion 
for the case of the normal propagation of radiation through the structure with 
the same dielectric permittivities of the substrate and cover layers. The simplifications listed do not change either the character of the high-order nonlinear response or the numerical estimates for the efficiency of up-conversion.

In summary, we have demonstrated that the heating by a strong ac field results in
an essentially nonharmonical response. This leads to the effective generation of
the third harmonic and opens a possibility for the efficient up-conversion
of mm pumping into THz signal. An experimental study of the process considered and
the electrodynamical calculations of a graphene structure integrated to an antenna
or into a THz resonator are very opportunely now. \\

The author is grateful to E. I. Karp for insightful comments.

\end{document}